\documentstyle[12pt,epsfig]{article}
\begin{document}                                                                
\begin{titlepage}                                                               
\load{\normalsize}{\rm}                                                         
\begin{flushright}
IFIC/96-15
\end{flushright}
\vspace{2.0cm}                                                                    
\begin{quote}                                                                   
\centering                                                                      
{\LARGE                                                                         
{\bf A comparative study on two characteristic  parametrizations 
for high energy pp and \=pp total cross sections}}
\end{quote}                                                                     
\begin{quote}                                                                   
\Large                                                                          
\vspace{1cm}                                                                    
\begin{center}                                                                  
A. Bueno$^1$ and J. Velasco$^2$                                                         
                                                                                
\large                                                                          
\it
$^1$ High Energy Physics Laboratory, Harvard University, Cambridge, 02138
MA, USA \\
                                                                             
$^2$IFIC, Centro Mixto Universitat                                                  
de Valencia-CSIC 46100-Burjassot, Valencia, Spain                            
\end{center}                                                                    
\end{quote}                                                                     
                                                                                
\vspace{0.5cm}                                                                    
\begin{quote}                                                                   
\rm                                                                             
\normalsize                                                                     
\begin{center}                                                                  
{\bf Abstract}\\[0.3cm]                                                                
\end{center}                                                                    
Available high energy data for both pp and \=pp total
cross sections  
($ 5 \ GeV \ < \ \sqrt s \ < \ 1.8 \ TeV$) are described by means
of two well-known distinct             
parametrizations, characteristic of  theoretical (``Regge-like" expression) and
experimental (``Froissart-Martin-like" expression) practices, respectively. 
Both are compared       
from the statistical point of view.                               
For the whole set of  present data  statistical analysis ($\chi^2/d.o.f.$) 
seems to favour a ``Froissart-like" ((ln $s$)$^{\gamma \approx 2}$ )
rise of the total cross section                 
rather than a ``Regge-like" ($s^{\epsilon}$) one. 
\end{quote}
\vspace*{1.0cm}             
{\em PACS number(s):} 13.85Lg, 13.85.-t \\[0.4cm]
\vspace*{0.5cm}
\begin{center}
{\em Submitted to Physics  Letters B}
\end{center}
\end{titlepage}                                                                 
\normalsize                                                                     
\pagebreak                                                                      
\indent                                                                         
\parindent 1.3cm                                                                
\marginparpush 5pt                                                              
\normalsize
\hspace*{0.73cm} It is a well-established fact 
that hadronic total cross sections rise with energy (\cite{ISR},               
\cite{ISR2}, \cite{UA4})                                                                     
but the actual energy dependence of this rise is still an open question.        
We dispose of many  parametrizations (see ref. \cite{land1}, \cite{amal}, 
\cite{para3}, \cite{para4}, which are by no means exhaustives) 
to describe the available data.
Although different energy                                           
behaviours are proposed, as a matter of  fact, 
most of them describe high energy data fairly           
well.                                                                                                                                                        
\par
In this paper  the explanation of  how total cross sections          
grow and the predictions for future colliders energies play a secondary        
role. Our aim                                                                   
is {\bf to compare  from a statistical point of view} two               
characteristic  parametrizations. 
The first  is a semi-empirical parametrization, 
based on Regge theory  
and asymptotic theorems, which           
assumes a (ln $s$)$^\gamma$ (``Froissart-like") behaviour
at high-energy. It has 
successively been used by experimentalists to describe their
data, from the ISR \cite{amal} to the S\=ppS \cite{extra}. Moreover
it has proved to be very successful in predicting the behaviour 
of the total cross section from the ISR to the S\=ppS energies,
in particular its considerable rising \cite{UA4}.
The other one, developed by Donnachie and Landshoff \cite{land1}, 
establishes             
that total cross sections grow as a power of the energy: 
$s^{\epsilon}$. Even if we know
that $\sigma^{tot}$ cannot ultimately grow as a power of the energy, 
because unitarity will break down,
it is claimed that it is perfectly valid in the non-asymptotic domain 
which has been explored up to now.
It has a sound theoretical basis on Regge theory and has successfully
been applied to describe  
a variety of hadronic processes \cite{land2}:
\=pp, pp, $\pi^\pm$p, $K ^\pm$p, 
$\gamma$p, \=pn, pn. It may well be considered as 
one of the most popular  from the theoretical point of view.
Its  simple and compact form make it a very useful expression
for representing a great variety of data as exemplified 
in the last version of 
the Review of Particle Properties  \cite{PDG}.
\par
In our choice we have been helped by  the recent measurement 
of the real part of the forward elastic scattering amplitude 
at the S\=ppS Collider        
\cite{rho}. The measurement has proved  that, 
for the energy scale of the present                  
accelerators, the contribution of the odd under crossing part of the            
scattering amplitude \cite{ode}, \cite{ode2} is negligible.                     
This fact make plausible   to omit         
all the parametrizations which include odderon effects.                      
\par
Our comparative study will be done for pp and \=pp processes only
as  they accumulate the most precise and energy-broader set of data. 
As far as the  other hadronic processes are concerned 
the comparison between the two                      
parametrizations in the high energy regime 
has much less interest due to the  absence          
of statistics in all the cases for $\sqrt{s} > 30$ GeV \cite{PDG}.
\vspace*{0.25cm} 
\par
Let us assume that total cross sections can be described           
by any of the two following expressions,
hereafter called $P_{1}$  and $P_{2}$, respectively (a detailed explanation
of their motivation  can  be found in ref. \cite{land1} for $P_{1}$, and 
ref. \cite{rho} for $P_{2}$):\\
                                                                                
\begin{equation}                                                                
\sigma_{\pm}^{tot} = A_1 \ E^{- N_1} \ \mp \ A_2 \ E^{-N_2} \ + \               
C_0 \ + \ C_2 \ [ln \ (s/s_0)]^\gamma                                
\end{equation}

\begin{equation}                                                                
\sigma_{\pm}^{tot} = X \ s^\epsilon \ + \ Y_{\pm} \ s^{-\eta}                   
\vspace*{0.6cm}                                                                 
\end{equation}                                                                  
where $+$($-$) stands for pp (\=pp) 
diffusion. $\sigma^{tot}$ is measured       
in mb and energy in GeV, being $E$ the energy measured in the lab frame.        
The scale factor  $s_0$ have been arbitrarily chosen equal to 1 GeV$^2$.       
\par
In $P_{1}$ the first two terms are Regge-type terms which describe
the behaviour at low energy and the difference between $\sigma_{\bar{p}p}$
and $\sigma_{pp}$. The remaining ones describe the high-energy
behaviour. In case $C_2 \leq \pi /m^2_{\pi}$ and $\gamma \leq 2$, 
this parametrization  is compatible with
the asymptotic Froissart-Martin bound \cite{Froissart}.
\par
In $P_{2}$ the first term arises from pomeron exchange and the second one from
$\rho, \omega, f,  a \ $ exchange. The coefficient $X$ is the same for
pp and \=pp. Both $\epsilon$ and $\eta$ are effective powers,
slowly varying with $s$. Previous theoretical work indicated that
$\epsilon$ should be close to 0.08 and that $\eta$ is about 1/2. 
>From their analysis
of pp and \=pp data, 
Donnachie and Landshoff conclude that $\epsilon$ and $\eta$
 can be treated as constants
with values $\epsilon = 0.0808$ 
and  $\eta =  0.4525$.
\vspace*{0.25cm}
\par
In order to compare  the approaches 
embodied in $P_{1}$
and $P_{2}$,   three different kinds of fits,  using  three different
data sets  for each fit, have been  performed. 
On each data set the first fit, $F_{1}$,  
is done using parametrization $P_{1}$  and  the second fit, $F_{2}$, 
is done with $P_{2}$.
In all the cases all the parameters are allowed to vary.
\par
Furthermore,  in order to make a faithful comparison with the results
obtained by Donnachie and Landshoff \cite{land2}
a third fit, $F_{3}$, is carried out. 
It is just an  $F_{2}$ 
fit where the parameters X, Y, $\epsilon$ and $\eta$ have been fixed 
to the values quoted by them.
\vspace*{0.25cm}
\par
For the first fit, the experimental data set consists in all available 
measurements of $\sigma^{tot}$ and $\rho$,
in the  energy domain which spans from $\sqrt{s} = 5$ GeV up          
to $\sqrt{s} = 546$ GeV. The existing                  
discrepancy between CDF (80.6 $\pm$ 2.3 mb) \cite{CDF} 
and E710 (72.8 $\pm$ 3.1 mb) \cite{E710} total cross             
section measurements at the Tevatron ($\sqrt{s} = 1.8$ TeV) 
leads us to exclude these          
points from the fits. This means that we are left with            
103 points.                    
\par                                     
The fits have been performed using the once-subtracted dispersion 
relations \cite{Soding}                  
\begin{equation}                                                                
\rho_{\pm} (E) \ \sigma_{\pm} (E) = {C_{s}\over p} +                            
{ E \over \pi p} \ \int_m^{\infty} \ dE^\prime \ p^\prime \                     
  \left[ {\sigma_{\pm} (E^\prime) \over E^\prime \                              
  (E^\prime \ - \ E)} \ - \ {\sigma_{\mp} (E^\prime) \over E^\prime \           
  (E^\prime \ + \ E)} \right]                                                   
\vspace*{0.6cm}                                                                 
\end{equation}                                                                  
where  $C_s$ is the substraction constant. It is a  {\em simultaneous} fit 
of $\sigma^{tot}$ and $\rho$, that is, 
we minimize the $\chi^2$ function\\[0.1in]
\begin{center}
$\chi^2$ = $ \chi^2_{\sigma_{\bar{p}p}} + \chi^2_{\rho_{\bar{p}p}} +
\chi^2_{\sigma_{pp}} + \chi^2_{\rho_{pp}} $ \\[0.2in]
\end{center}
 The method has proved to be very succesful in the past 
 \cite{amal} predicting the observed rise of $\sigma^{tot}$ from the ISR to 
the S\=ppS.  Recently it has been used, with the last data available, 
to provide
predictions on the behaviour of $\sigma^{tot}$ up to $\sqrt{s} = 100 $ TeV
\cite{extra}.
A detailed study of the technique can be found in \cite{Tesis}.
\par  
In $F_{1}$ we are left with eight free parameters                         
($A_1$, $N_1$, $A_2$, $N_2$, $C_0$, $C_2$, $\gamma$, $C_s$). 
For the $F_{2}$   
we have six free parameters ($X$, $Y_+$, $Y_-$,            
$\eta$, $\epsilon$, $C_s$). 
Finally in $F_{3}$
we have only one free parameter                
($C_s$).            
\par
Table 1 shows the values of the parameters obtained 
for the best $F_{1}$, $F_{2}$ and $F_{3}$ fits. Table 2 quotes,      
for these best fits,  the values of $\chi^2$ and  $\chi^2/d.o.f.$
We have included also 
the predicted numerical values for          
$\sigma^{tot}$ for each of the three considered cases at 
the  energies of the  S\=ppS, Tevatron and LHC.  In table 3 
the experimental values of $\sigma^{tot}$ at the S\=ppS 
and Tevatron are compared to the predictions of $F_{1}$, $F_{2}$, and
$F_{3}$.
\par
It clearly 
appears  that $F_{1}$, with the  dominant (ln $s$)$^{\gamma        
\approx 2}$ term at high energies  ($\chi^2/d.o.f.$ = 0.8) gives 
 by far the best result. It perfectly matches the S\=ppS 
measured value. It is interesting to observe that  
its prediction at the Tevatron  lies right 
at the middle $(76.5 \pm 2.3$ mb)
of the conflicting claims for $\sigma^{tot}$ 
of both  Tevatron experiments, CDF and E710.
As for  $F_{2}$, even if it may be
considered as an acceptable fit  ($\chi^2/d.o.f.$ = 1.6)
it has, at least, two difficulties:  on the one hand,
the values obtained for $\epsilon$, $0.0644 \pm 0.0015$, 
and $\eta$, $0.5433 \pm 0.0075$, represent a challenge for the model. 
They are far away 
from their theoretical estimations quoted previously.
And, on the other hand, 
  the numerical         
values obtained for $\sigma^{tot}$ (see table 2), being  
much lower than the
measured ones, rule it out. 
 Finally
$F_{3}$   is clearly {\it statistically} ruled out  
($\chi^2/d.o.f.$ = 4.5) with respect to the others fits, although 
its $\sigma^{tot}$ predictions might appear acceptable. 
Indeed, Donnachie and Landshoff
considered a definite  success of their parametrization 
the prediction, in 1985 \cite{dl2}, of  about 73 mb for $\sigma^{tot}$ 
at the Tevatron. As we have said, the accepted value of $\sigma^{tot}$
at 1.8 TeV is far from clear up to now. 
Waiting for the measurement of $\sigma^{tot}$ at the LHC 
we find this parametrization 
statistically unsupported by present data.
\vspace*{0.25cm}
\par
In our second step,  the fits are less powerful: we   
restrict them    to 
the $\sigma^{tot}_{pp}$ and $\sigma^{tot}_{\bar{p}p}$ 
experimental data sample (69 points), 
excluding the $\rho_{{\bar{p}p}, pp}$ measurements, which is
the usual practice.
Now we have seven  
free parameters for $F_{1}$
 and  five for $F_{2}$.        
In this case, $F_{3}$
 is not a real fit 
because we do not minimize $\chi^2$.       
\par
 Table 4 shows              
the parameters as given by the best fits and table 5 gives
in addition         
to $\chi^2$ and $\chi^2/d.o.f.$, the values                          
for $\sigma^{tot}$ at 
the  energies of the S\=ppS, Tevatron and LHC.  
\par
In 
figure 1 we have depicted            
the results of these three fits together with the 
experimental data on $\sigma^{tot}$. Recent cosmic rays results 
\cite{honda} from the Akeno Observatory obtained from the analysis of 
proton-air interactions at ultrahigh energies are also plotted.
\par
The results show no sensible changes with
respect to the ones of the 
previous step and the same discussion applies.   
Again $F_{1}$ is strongly supported by the data,
which is not the case for
$F_{2}$. $F_{3}$ yields, once again, $\sigma^{tot}$ predictions consistent with
current experimental evidence, but its large $\chi^2/d.o.f.$ value
rules it out from the statistical point of view.
\vspace*{0.25cm}
\par                                                    
Finally we investigate how the variation of the energy domain influences        
the results previously obtained.  
We restricted the fits to a smaller energy domain,                 
in the sense of increasing energy, $10 \leq \sqrt{s} \leq 546$ GeV, 
but still keeping enough data 
(42 points) to make results sensible.                                     
\par
The trend previously observed are, in spite of the bigger 
statistical errors due to the smaller number of data points, not altered:
the results for the different fits are practically the same
as an inspection of tables 6 and 7 shows.           
$F_{1}$
 always gives                              
the lowest $\chi^2/d.o.f.$ value. $F_{2}$
 does not reproduce
the experimental points and $F_{3}$
, which does better than $F_{2}$, has the worst
$\chi^2$ of all three fits.
\vspace*{0.25cm}
\par
In conclusion, 
although it is claimed  that  (ln $s$)$^{\gamma \approx        
2}$ and $s^{0.0808}$ rises adequately reproduce  
high-energy pp
and \=pp data,          
from a careful analysis of
present experimental evidence,
we have shown that statistically a (ln $s$)$^{\gamma \approx 2}$ fit is         
strongly favoured. Better $\chi^2/d.o.f.$ may be  obtained, in the
$s^{\epsilon}$ model, at the price of  lower                
$\epsilon$ values, but for these  cases the resulting $\sigma^{tot}$            
values are clearly unrealistic.                                 
\vspace*{0.4cm}
\begin{center}
\large {\bf
Acknowledgements} \\[0.4cm]
\end{center}
\normalsize
This research has been supported by  CICYT grant number  AEN93-0792. 

\newpage 
\vspace*{4cm}
\begin{center}
{\bf Figure captions}
\end{center}
\vspace*{2cm}
{\em Fig. 1}  Total cross sections data from accelerators 
and from cosmic rays are shown together with 
the best $F_{1}$, $F_{2}$ and $F_{3}$ fits using  $\sigma^{tot}_
{pp}$ and $\sigma^{tot}_{\bar{p}p}$ data.
\newpage
..
\newpage
\begin{table}[t]
\begin{center}
\begin{tabular}{|c|c|c|c|c|} \hline
 & \multicolumn{4}{c|}{ } \\
Fit type & \multicolumn{4}{c|}{Parameters} \\
 & \multicolumn{4}{c|}{ } \\ \hline
 & & & & \\
 (ln $s$)$^\gamma$ & $A_1 = 42.5^{+ 2.0}_{- 1.6} $ & 
$N_1 = 0.45^{+ 0.08}_{- 0.06}$
& $A_2 = 25.5^{+ 0.5}_{- 0.4}$ &
 $N_2 = 0.565^{+ 0.005}_{- 0.004}$ \\
 & & & & \\ \cline{2-5}
 & & & & \\
 & $C_0 = 30.0^{+ 3.0}_{- 4.0}$ & $C_2 = 0.10^{+ 0.15}_{- 0.06}$ &
$\gamma = 2.25^{+ 0.35}_{- 0.31}$ & $C_s$ = $- 57.0 \pm 4.0$  \\
 & & & & \\ \hline
 & & & & \\
Regge 1 & $Y_+ = 53.90 \pm 0.1$ & $Y_- = 121.15 \pm 2.20$ &
$\eta = 0.5433 \pm 0.0075$ & \\
 & & & & \\ \cline{2-4}
 & & & & \\
 & $X = 25.16 \pm 0.29$ & $\epsilon = 0.0644 \pm 0.0015$ &
$C_s$ = $- 37.7 \pm 0.7$ & \\
 & & & & \\ \hline
 & & & & \\
Regge 2 & $Y_+$ = 56.08 & $Y_-$ = 98.39 & $\eta$ = 0.4525 & \\
 & & & & \\ \cline{2-4}
 & & & & \\
 & $X$ = 21.70 & $\epsilon$ = 0.0808 & $C_s$ = $- 4 \pm 17$ & \\
 & & & & \\ \hline
\end{tabular}
\end{center}
\caption{Values of the parameters given by the best dispersion relations
fits}
\vspace*{10cm}
\end{table}
\newpage
\begin{table}[t]
\begin{center}
\begin{tabular}{|c|c|c|c|c|c|} \hline
 & & & & & \\
Fit type & $\chi^2$ & $\chi^2/d.o.f.$ & $\sigma^{tot}$ (546 GeV) &
$\sigma^{tot}$ (1.8 TeV) & $\sigma^{tot}$ (14 TeV) \\
 & & & & & \\ \hline
 & & & & & \\
(ln $s$)$^\gamma$ & 78.5 & 0.8 & 61.8 $\pm$ 0.7& 76.5 $\pm$ 2.3& 110. $\pm$ 8\\
 & & & & & \\ \hline
 & & & & & \\
Regge 1 & 153.9 & 1.6 & 56.8 $\pm$ 0.4 & 66.1 $\pm$ 0.7 & 86.0 $\pm$ 1.4 \\
 & & & & & \\ \hline
 & & & & & \\
Regge 2 & 456.3 & 4.5 & 60.4 & 73.0 & 101.5 \\
 & & & & & \\ \hline
\end{tabular}
\end{center}
\caption{$\chi^2$ and $\sigma^{tot}$ values obtained by fitting
with dispersion relations the available $\sigma^{tot}$ and $\rho$
experimental data for pp and \=pp scattering. $\sigma^{tot}$ is measured in mb.}
\vspace*{10.cm}
\end{table}
\begin{table}[t]
\begin{center}
\begin{tabular}{|c|c|c||c|c|c|} \hline
 & & & & & \\
$ \sqrt s $ (TeV) & Data & $ \sigma^{tot}$ (mb) & $F_{1}$ & $F_{2}$& $F_{3}$  \\
 & & & & & \\ \hline
 & & & & &  \\
0.55 & UA4 & $62.2 \pm 1.5$ &  $61.8 \pm 0.7$ & $ 56.8 \pm 0.4$ & 60.4\\
 & CDF & $61.5 \pm 1.0$ & & &\\
 & & & & &\\ \hline
 & & & & & \\
1.8 & E710& $72.8 \pm 3.1$ & $76.5 \pm 2.3$ & $ 66.1 \pm 0.7 $& 73.0\\
 & CDF & $80.6 \pm 2.3$ & & &\\
 & & & & &\\ \hline
\end{tabular}
\end{center}
\caption{ Experimental $\sigma^{tot}$ values in mb at the S\=ppS (0.55 TeV)
and Tevatron (1.8 TeV) and best $ F_{1}$, $F_{2}$ and $F_{3}$ predictions.}
\vspace*{10.cm}
\end{table}
\newpage

\begin{table}[t]
\begin{center}
\begin{tabular}{|c|c|c|c|c|} \hline
 & \multicolumn{4}{c|}{ } \\
Fit type & \multicolumn{4}{c|}{Parameters} \\
 & \multicolumn{4}{c|}{ } \\ \hline
 & & & & \\
 (ln $s$)$^\gamma$ & $A_1 = 42.5^{+ 2.7}_{- 2.0}$ 
& $N_1 = 0.49^{+ 0.08}_{- 0.11}$
& $A_2 = 25.2^{+ 0.1}_{- 0.2} $ &
 $N_2 = 0.562^{+ 0.003}_{- 0.002} $ \\
 & & & & \\ \cline{2-5}
 & & & & \\
 & $C_0 = 32.4^{+ 2.7}_{- 6.0}$ & $C_2 = 0.063^{+ 0.280}_{- 0.040}$ &
$\gamma = 2.42^{+ 0.4}_{- 0.6}$ &  \\
 & & & & \\ \hline
 & & & & \\
Regge 1 & $Y_+ = 51.52 \pm 0.15$ & $Y_- = 118.71 \pm 2.2$ &
$\eta = 0.5431 \pm 0.0075$ & \\
 & & & & \\ \cline{2-4}
 & & & & \\
 & $X = 25.62 \pm 0.30$ & $\epsilon = 0.0618 \pm 0.0015$ &  & \\
 & & & & \\ \hline
 & & & & \\
Regge 2 & $Y_+$ = 56.08 & $Y_-$ = 98.39 & $\eta$ = 0.4525 & \\
 & & & & \\ \cline{2-4}
 & & & & \\
 & $X$ = 21.70 & $\epsilon$ = 0.0808 &  & \\
 & & & & \\ \hline
\end{tabular}
\end{center}
\caption{Values of the parameters obtained after fitting $\sigma^{tot}_
{pp}$ and $\sigma^{tot}_{\bar{p}p}$ data. Tevatron
measurements are excluded from the fit. The lowest energy limit
corresponds to 5 GeV.}
\vspace*{10.cm}
\end{table}
\newpage 

\begin{table}[t]
\begin{center}
\begin{tabular}{|c|c|c|c|c|c|} \hline
 & & & & & \\
Fit type & $\chi^2$ & $\chi^2/d.o.f.$ & $\sigma^{tot}$ (546 GeV) &
$\sigma^{tot}$ (1.8 TeV) & $\sigma^{tot}$ (14 TeV) \\
 & & & & & \\ \hline
 & & & & & \\
(ln $s$)$^\gamma$ & 43.1 & 0.7 & 61.7 $\pm$ 1.3 & 76.7 $\pm$ 4.0&
112. $\pm$ 13 \\
 & & & & & \\ \hline
 & & & & & \\
Regge 1 & 87.0 & 1.4 & 55.9 $\pm$ 0.4 & 64.7 $\pm$ 0.6 & 83.4 $\pm$ 1.4 \\
 & & & & & \\ \hline
 & & & & & \\
Regge 2 & 395.3 & 5.7 & 60.4 & 73.0 & 101.5 \\
 & & & & & \\ \hline
\end{tabular}
\end{center}
\caption{$\chi^2$ and $\sigma^{tot}$ (in mb) values obtained by fitting
available data on pp and \=pp total cross sections. The energy range
spans from 5 GeV to 546 GeV.}
\vspace*{10.cm}
\end{table}
\newpage

\begin{table}[t]
\begin{center}
\begin{tabular}{|c|c|c|c|c|} \hline
 & \multicolumn{4}{c|}{ } \\
Fit type & \multicolumn{4}{c|}{Parameters} \\
 & \multicolumn{4}{c|}{ } \\ \hline
 & & & & \\
 (ln $s$)$^\gamma$ & $A_1 = 49.3^{+ 6.4}_{- 5.8}$ & $N_1 = 0.59^{+ 0.16}_{-
0.20}$ &
$A_2 = 25.4^{+ 0.2}_{- 0.3}$ &
 $N_2 = 0.562^{+ 0.002}_{- 0.001}$ \\
 & & & & \\ \cline{2-5}
 & & & & \\
 & $C_0 = 34.4^{+ 3.1}_{- 7.3}$ & $C_2 = 0.034^{+ 0.220}_{- 0.060}$ &
$\gamma = 2.64^{+ 0.50}_{- 0.32}$ &  \\
 & & & & \\ \hline
 & & & & \\
Regge 1 & $Y_+$ = 60.87 $\pm$ 1.30 & $Y_-$ = 124.41 $\pm$ 7.30 &
$\eta$ = 0.5362 $\pm$ 0.0190 & \\
 & & & & \\ \cline{2-4}
 & & & & \\
 & $X$ = 24.26 $\pm$ 0.50 & $\epsilon$ = 0.0687 $\pm$ 0.0024 &  & \\
 & & & & \\ \hline
 & & & & \\
Regge 2 & $Y_+$ = 56.08 & $Y_-$ = 98.39 & $\eta$ = 0.4525 & \\
 & & & & \\ \cline{2-4}
 & & & & \\
 & $X$ = 21.70 & $\epsilon$ = 0.0808 &  & \\
 & & & & \\ \hline
\end{tabular}
\end{center}
\caption{Same as table 4, but now the lowest limit of the energy interval
corresponds to 10 GeV.}
\vspace*{10.cm}
\end{table}
\newpage 

\begin{table}[t]
\begin{center}
\begin{tabular}{|c|c|c|c|c|c|} \hline
 & & & & & \\
Fit type & $\chi^2$ & $\chi^2/d.o.f.$ & $\sigma^{tot}$ (546 GeV) &
$\sigma^{tot}$ (1.8 TeV) & $\sigma^{tot}$ (14 TeV) \\
 & & & & & \\ \hline
 & & & & & \\
(ln $s$)$^\gamma$ & 28.5 & 0.8 & 61.7 $\pm$ 1.3 & 77.4 $\pm$ 4.2 & 116. $\pm$
16\\
 & & & & & \\ \hline
 & & & & & \\
Regge 1 & 39.5 & 1.1 & 57.8 $\pm$ 0.6 & 68.0 $\pm$ 1.0 & 90.0 $\pm$ 2.3 \\
 & & & & & \\ \hline
 & & & & & \\
Regge 2 & 103.4 & 2.5 & 60.4 & 73.0 & 101.5 \\
 & & & & & \\ \hline
\end{tabular}
\end{center}
\caption{Same as table 5, but now the lowest limit of the energy
interval corresponds to 10 GeV.}
\end{table}

\begin{figure}
\begin{center}
\mbox{\epsfig{file=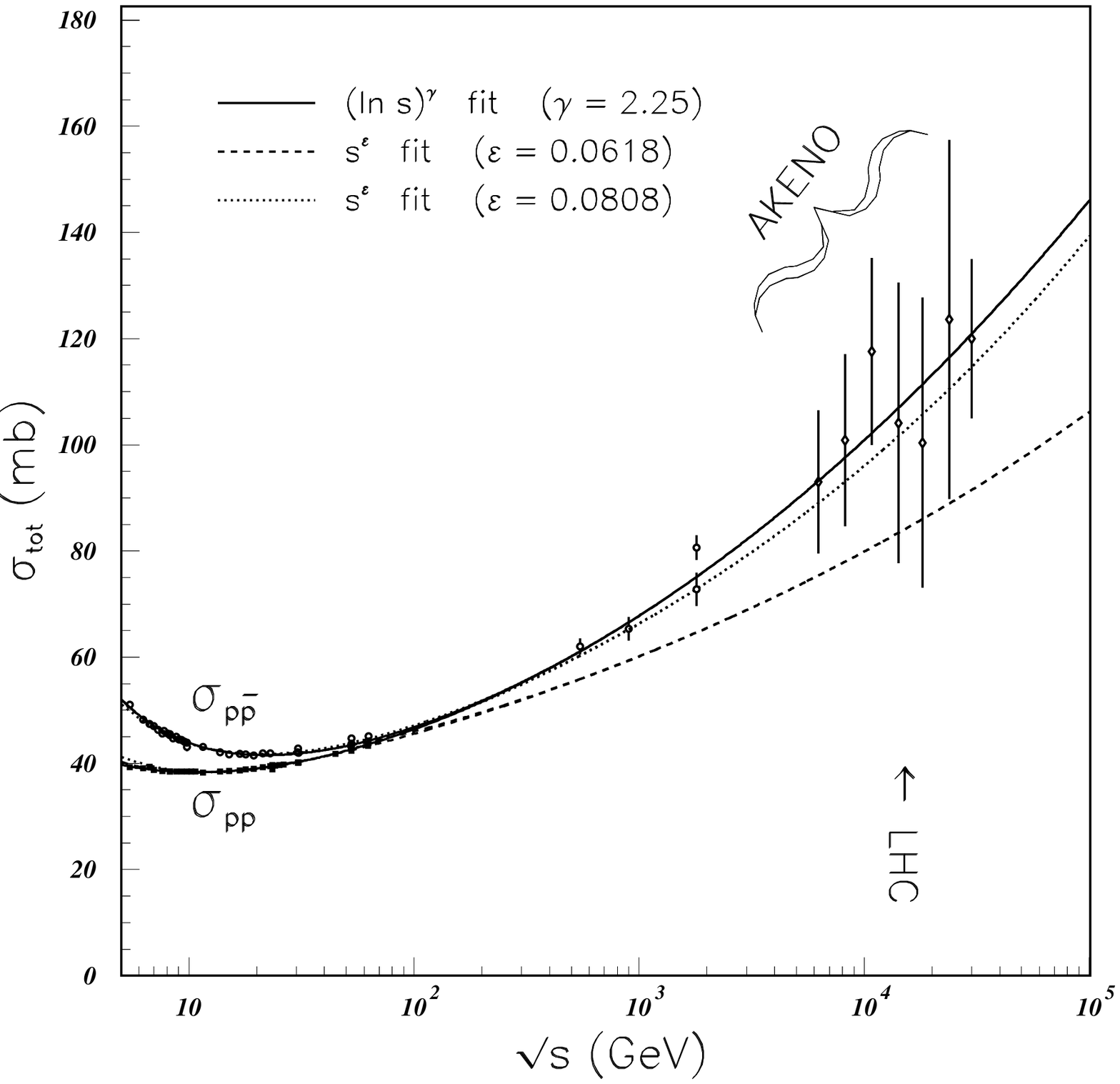,height=15.cm}}
\end{center}
\end{figure}

\end{document}